\documentclass[journal,onecolumn,12pt]{IEEEtran}
\usepackage{geometry}
\usepackage{pgf}
\usepackage[square, comma, sort&compress, numbers]{natbib}
\usepackage[english]{babel}
\usepackage[utf8]{inputenc}
\usepackage{booktabs}
\usepackage[T1]{fontenc}
\usepackage{algorithm}
\usepackage{algorithmic}
\usepackage{datetime}
\usepackage{verbatim}
\usepackage{array}
\usepackage{xspace}

\usepackage{amsmath,amsfonts,amssymb,amsopn,amsthm,amscd}
\usepackage{mathtools}
\theoremstyle{plain}
\usepackage{graphicx} %use graph format
\usepackage{enumerate}
\newcommand{\Ff}{{\mathbb F}}

\newcommand\cc{{\mathcal C}}        %

\def\Tr{\operatorname{Tr}}

\newtheorem{thm}{Theorem}%[section]
\newtheorem{lem}[thm]{Lemma}

\newtheorem{prop}[thm]{Proposition}

\newtheorem{example}[thm]{Example}
\newtheorem{remark}[thm]{Remark}
\def\Tr{\operatorname{Tr}}

\floatname{algorithm}{Construction}

\usepackage[colorlinks=true]{hyperref}
\usepackage{enumitem}

\usepackage[colorlinks=true]{hyperref}
\usepackage{enumitem}

\newcommand{\QEDB}{\hfill \ensuremath{\Box}}

\begin{document}
\title{New Interleaving Constructions of Asymptotically Optimal Periodic Quasi-Complementary Sequence Sets}
\author{\centerline{Gaojun Luo, Martianus Frederic Ezerman, and San Ling}
\thanks{G. Luo, M. F. Ezerman, and S. Ling are with the School of Physical and Mathematical Sciences, Nanyang Technological University, 21 Nanyang Link, Singapore 637371, e-mails: $\{\rm gaojun.luo, fredezerman, lingsan\}$@ntu.edu.sg.}
\thanks{The authors are supported by Nanyang Technological University Research Grant No. 04INS000047C230GRT01.}
\thanks{This work has been submitted to the IEEE for possible publication. Copyright may be transferred without notice, after which this version may no longer be accessible.}
%\thanks{Copyright (c) 2021 IEEE. Personal use of this material is permitted. However, permission to use this material for any other purposes must be obtained from the IEEE by sending a request to pubs-permissions@ieee.org.}
}
\maketitle

\begin{abstract}
The correlation properties of sequences form a focal point in the design of multiple access systems of communications. Such a system must be able to serve a number of simultaneous users while keeping interference low. A popular choice for the set of sequences to deploy is the quasi-complementary sequence set (QCSS). Its large set size enables the system to accommodate a lot of users. The set has low nontrivial correlation magnitudes within a zone around the origin. This keeps undue interference among users under control. A QCSS performs better than the perfect complementary sequence set (PCSS) does in schemes with fractional delays.

The optimality of a set of periodic sequences is measured by its maximum periodic correlation magnitude, for which there is an established lower bound to aim at. For a fixed period, optimal sets are known only for very restricted parameters. Efforts have therefore been centered around the constructions of asymptotically optimal sets. Their periods are allowed to be as large as sufficient to establish optimality. In this paper we share an insight that a sequence set that asymptotically attains the Welch bound generates an asymptotically optimal periodic QCSS by interleaving. One can simply use known families of such sequence sets to construct the desired QCSSs. Seven families of QCSSs with specific parameters are shown as examples of this general construction. We build upon the insight to propose two new direct constructions of asymptotically optimal QCSSs with very flexible parameters without interleaving. The flexibility enhances their appeal for practical implementation. The mathematical tools come from the theory of groups in the form of additive and multiplicative characters of finite fields.
\end{abstract}

\begin{IEEEkeywords}
Quasi-complementary sequence set (QCSS), asymptotically optimal, periodic correlation, polyphase sequence set, multi-carrier code-division multiple-access (MC-CDMA)
\end{IEEEkeywords}

\section{Introduction}\label{sec:intro}

Numerous contemporary communications systems support multiple users sharing the same frequency band simultaneously by multiplexing. To prevent undue interference among users, such a multiple access system employs a special coding scheme. The chosen scheme typically consists of a set of sequences with desirable properties that enhance both efficiency and reliability. Two important measures form the overall correlation property of the set. The first is the auto-correlation which measures how each sequence correlates to its time-shifted self. The second is the cross-correlation which examines how each sequence correlates to any other sequence within the set.

A ubiquitous multiple access system is the multi-carrier code-division multiple-access (MC-CDMA). It spreads signals over parallel (sub)carriers by assigning to every user a two-dimensional matrix. It transfers the user's data symbol by transmitting all of the row sequences simultaneously over the carriers \cite{Chen1,Chen2,Chen3}. The rows of the matrix associated to a user can be seen as sequences. This way, the matrix is a set of sequences. A natural question is how to design a good set of sequences. There are, in fact, many choices already available in the literature.

One such choice is the perfect complementary sequence set (PCSS). This is a collection of two-dimensional matrices in which nontrivial auto-correlations and cross-correlations of the row sequences sum to zero for any nonzero time-shift. A PCSS, however, has a significant drawback. Its set size, which is the number of users that the system can serve, cannot exceed the number of row sequences in the two-dimensional matrices. If a large number of users need to be supported, then PCSSs become less efficacious to be deployed in an MC-CDMA. To support more users than a PCSS can, two generalizations have been proposed.

The first one is referred to as a {\it low-correlation zone complementary sequence set} (LCZ-CSS). Like a PCSS, an LCZ-CSS also consists of two-dimensional matrices. Unlike the PCSS, it possesses low nontrivial correlation magnitudes \emph{within a zone around the origin} \cite{ZCZ1,ZCZ2,LCZ}. If the correlation magnitudes are all equal to zero within the zone, then the set is called a {\it zero correlation zone complementary sequence set} (ZCZ-CSS).

Liu, Parampalli, Guan, and Boztas in \cite{Liu1}, followed by Liu, Guan, and Mow in \cite{Liu2}, introduced and studied the properties of the second extension of PCSS, which is called a {\it quasi-complementary sequence set} (QCSS). A QCSS contains two-dimensional matrices such that the maximum nontrivial correlation sum is small but nonzero for any nonzero time-shift. The large set sizes and low interference performance of QCSSs, established in \cite{Liu2,Liu3}, make them desirable, attracting scholars to carry out extensive studies on their properties. Some transmission schemes have delays of non-integral multiple of half chip duration \cite{Liu4}. In such a ``fractional-delay'' scheme, QCSSs typically perform better than PCSSs do \cite{Samad}.

\subsection{Periodic QCSSs and polyphase sequence sets}\label{subsec:QCSS}

Given an $M \in \mathbb{N}$, let $[M]:=\{0,1,\cdots,M-1\}$. For each $i \in[M]$, we use $\mathcal{C}^i$ to denote the $K \times N$ matrix
\[
\mathcal{C}^i =
\begin{pmatrix}
\mathbf{c}_0^i\\
\mathbf{c}_1^i\\
\vdots\\
\mathbf{c}_{K-1}^i
\end{pmatrix} =
\begin{pmatrix}
c_0^i(0) & c_0^i(1) & \cdots & c_0^i(N-1)\\
c_1^i(0) & c_1^i(1) & \cdots & c_1^i(N-1)\\
\vdots&\vdots &\ddots &\vdots\\
c_{K-1}^i(0) & c_{K-1}^i(1) & \cdots & c_{K-1}^i(N-1)\\
\end{pmatrix},
\]
with the $k^{\rm th}$ constituent sequence of period $N$, for each $k \in [K]$, being written as
\[
\mathbf{c}_k^i := \left(c_k^i(0), c_k^i(1), \ldots, c_k^i(N-1)\right).
\]
Let $a^*$ denote the complex conjugate of $a$. Each entry of $\mathcal{C}^i$ is a complex number of modulus $1$ taken from a set $\mathcal{A}$. The {\it periodic correlation function} (PCF) between $\mathcal{C}^i$ and $\mathcal{C}^j$ is defined by
\begin{equation}\label{eq:PCF}
R_{\mathcal{C}^i,\mathcal{C}^j}(\tau) := \sum_{k=0}^{K-1} \sum_{t=0}^{N-1}c_k^i(t)c_k^{j*}(t+\tau) \mbox{, for each } \tau \in [N],
\end{equation}
with the addition $(t+\tau)$ in the argument done modulo $N$. Let $\mathfrak{C} := \{\mathcal{C}^0, \mathcal{C}^1, \ldots, \mathcal{C}^{M-1}\}$. The {\it maximum periodic correlation magnitude}, also called the {\it periodic tolerance} of $\mathfrak{C}$, is
\begin{equation}\label{eq:maxcor}
\vartheta_{\rm max} := \max_{i,j\in[M]} \left\{\left|R_{\mathcal{C}^i,\mathcal{C}^j}(\tau)\right| \mbox{ with } \tau\in[N] \mbox{ when } i\neq j \mbox{ and } \tau\in[N]\setminus\{0\} \mbox{ when } i=j\right\}.
\end{equation}
As a generalization of the Welch bound in \cite[Corollary (Periodic)]{Welch}, Liu {\it et al}. in \cite{Liu1} proved the following lower bound
\begin{equation}\label{bound}
\vartheta_{\rm max}\geq \vartheta_{\rm opt}=KN \sqrt{\frac{\frac{M}{K}-1}{MN-1}}.
\end{equation}
The bound reduces to the Welch bound when $K=1$. The set $\mathfrak{C}$ is called a {\it periodic sequence set} or a {\it signal set} with parameters $(N,M,\vartheta_{\rm max})$ if $K=1$. If $M>K>1$, then the set $\mathfrak{C}$ is a {\it periodic QCSS} over the alphabet $\mathcal{A}$ with {\it set size} $|\mathfrak{C}|=M$, {\it flock size} $K$, sequence {\it length} $N$, and {\it maximum periodic correlation magnitude} $\vartheta_{\rm max}$. We call such a set $\mathfrak{C}$ an $(M,K,N,\vartheta_{\rm max})$-QCSS. Each element of a periodic QCSS is called a {\it periodic complementary sequence}.

The maximum possible amount of multipath interference and multiuser interference is determined by $\vartheta_{\rm max}$. In practice, given $K$ and $N$, it is desirable to construct an $(M,K,N,\vartheta_{\rm max})$-QCSS with $M$ as large as possible and $\vartheta_{\rm max}$ as small as it can be. A QCSS is {\it optimal} if its $\vartheta_{\rm max}$ achieves the bound in (\ref{bound}). If this takes place only for sufficiently large $N$, then the periodic QCSS is said to be {\it asymptotically optimal}.

\subsection{Known results}\label{subsec:known}

We briefly outline existing main approaches. Thus far, optimal periodic QCSSs exist only for very restrictive parameters. Constructing any infinite family of optimal periodic QCSSs still stands as the major challenge. Despite our best effort as a community, the problem remains open. Most researchers have turned their focus into constructing asymptotically optimal periodic QCSSs.

Liu {\it et al}. proposed a framework to build periodic QCSSs in \cite{Liu1}. Based on the framework, a family of asymptotically optimal periodic QCSSs was designed from Singer difference sets. Along this line, Li, Yan, and Lv in \cite{Li} constructed a family of periodic QCSSs by using almost difference sets, instead of Singer's. In \cite{Li1}, Li, Liu, and Xu reported a new method to construct periodic QCSSs that generalizes the approach used in \cite{Liu1}. In two consecutive papers \cite{Li2,Li3}, the three authors of \cite{Li1} together with L. Tian, used characters over finite fields as powerful construction tools to build several families of asymptotically optimal QCSSs. The first paper relies on multiplicative characters whereas the second combines additive and multiplicative characters. The deployment of additive characters and multiplicative characters requires the constructions to be done over finite fields with large alphabet sizes. Very recently, Luo, Cao, Shi, and Helleseth in \cite{Luo1} overcame this shortcoming. They proposed asymptotically optimal QCSSs with small alphabet sizes using additive characters.

\subsection{Our contributions and techniques}\label{subsec:contrib}

The main contribution of this paper comes in the form of several constructions of asymptotically optimal periodic QCSSs with flexible parameters. The results and techniques can be summarized as follows.
\begin{enumerate}
\item The interleaving technique, introduced by Gong in \cite{Gong}, is instrumental in designing sequences with good correlation properties. Examples of sequences designed by this technique can be found in \cite{Chung1,
    HuH, Zhou}. In this paper, we use interleaving to devise a general construction of QCSSs. To be more precise, we generate an $(M,K,N, \vartheta_{\rm max})$-QCSS with $KN = n$ and $\vartheta_{\rm max} \leq \theta_{\max}$ for any periodic sequence set with parameters $(n,M,\theta_{\max})$ by the general framework. We further demonstrate that such a QCSS is asymptotically optimal with respect to the bound in (\ref{bound}), if the corresponding sequence set asymptotically meets the Welch bound.
\item We derive seven infinite families of asymptotically optimal QCSSs. They are listed as Entries $10$ to $16$ in Table \ref{tabl1}. The table makes it clear that these new QCSSs outperform previously known QCSSs in terms of the alphabet size and on the flexible choices of the flock size and length.
\item Inspired by the general construction of QCSSs from interleaving, we offer two more constructions of asymptotically optimal QCSSs. These direct constructions are listed as Entries $17$ and $18$ in Table \ref{tabl1}. The resulting two families of QCSSs have very flexible parameters. These are, respectively, $((\mu-1)N,N,N,N)$ and $((\mu-1)N,N-1,N,N)$, where $ 1 < N \in \mathbb{N}$ is odd and $\mu$ is the smallest prime factor of $N$. For a fixed length, the alphabet sizes of these QCSSs are smaller than those in \cite{Li3,Liu1}. As the requirements of suitable MC-CDMA systems vary across deployment scenarios, flexibility in the choices of feasible parameters greatly enhances the appeal of our new QCSSs.
\end{enumerate}

\begin{table}
\renewcommand{\arraystretch}{1.2}
\caption{The parameters of asymptotically optimal periodic QCSSs with $n>1$, set size $M$, flock size $K$, length $N$, and alphabet size $|\mathcal{A}|$.}
\label{tabl1}
\renewcommand{\arraystretch}{1.2}
\centering
\begin{tabular}{cccccccc}
\toprule
No. & $M$ & $K$ & $N$ & $\theta_{\rm max}$ & $|\mathcal{A}|$ &  Constraints  & Reference \\
\midrule
$1$ & $2^n-1$      &  $2^{n-1}-1$   & $2^n$      & $(2^n+2^{n/2})/2$& $2(2^n-1)$ & $n>1$                &\cite{Liu1}  \\
\hline
$2$ & $p$ &  $(p-1)/2$ & $p$ & $\leq(p+\sqrt{p})/2$& $p$ & $p$ is prime &\cite{Li1}  \\
\hline
$3$ & $p^n-2$ &  $(p^n-1)/2$ & $p^n-1$ & $\leq(\ell+3)/2$& $p^n-1$ & $p$ is odd prime; &\cite{Li2}  \\
& & & & & & $\ell=p^n+4\sqrt{p^n}$\\

$4$ & $p^{2n}-2$ &  $p^n$ & $p^{2n}-1$ & $p^n(p^{2n}+3)$& $p^{2n}-1$ & $p$ is prime  &  \\
\hline

$5$ & $p^n-1$ &  $(p^n-1)/2$ & $p^n-1$ & $\leq(p^n+\sqrt{p^n})/2$& $p(p^n-1)$& $p$ is odd prime  &\cite{Li3}  \\

$6$ & $p^n-1$ &  $p^{n-1}$ & $p^n-1$ & $\leq p^{n-\frac{1}{2}}$ & $p(p^n-1)$ &  $p$ is prime   &  \\

$7$ & $p^{2n}-1$ &  $p^n$ & $p^{2n}-1$ & $p^{3n/2}$& $p(p^{2n}-1)$ & $p$ is prime  & \\

$8$ & $2^n-1$ &  $2^{n-1}-1$ & $2^n-1$ & $2^{n-1}$& $2(2^n-1)$ & $n>1$  & \\
\hline

$9$ & $p^n$ &  $(p^n\pm s)/2$ & $p^n-1$ & $(p^n+s)/2$ & $p$&  $p$ is prime    & \cite{Luo1}   \\
& & & & & & and $s=o(p^n)$\\
\hline

$10$ & $p^n$ &  $K$ & $(p^n-1)/K$ & $\leq p^{\frac{n}{2}}+1$ & $p$ &  $p$ is odd prime  & Thm. \ref{thm3.2}   \\
     & & & & & &   and $K=o(p^n)$\\

$11$ & $2^{\frac{n}{2}}+1$ &  $K$ & $(2^n-1)/K$ & $\leq 2^{\frac{n}{2}}+1$ & $2$ &  $K=o(2^{\frac{n}{2}}+1)$   & \\

$12$ & $p^{\frac{n}{2}}$ &  $K$ & $(p^n-1)/K$ & $\leq p^{\frac{n}{2}}+1$ & $p$ &  $p$ is odd prime  &  \\
 & & & & & &   and $K=o(p^{\frac{n}{2}})$\\

$13$ & $2^{n}+1$ &  $K$ & $(2^n-1)/K$ & $\leq 2^{\frac{n}{2}}+1$ & $4$ &  $K=o(2^{n}+1)$   & \\

$14$ & $2^{n}$ &  $K$ & $(2^{n+1}-2)/K$ & $\leq 2^{\frac{n+1}{2}}+2$ & $4$ & $K=o(2^{n})$ & \\

$15$ & $p$ &  $K$ & $(p^2-p)/K$ & $\leq p$ & $p$ &  $p$ is odd prime  & \\
             & & & & & &  and $K=o(p)$\\

$16$ & $r$ &  $K$ & $(p^n-1)/K$ & $\leq p^{\frac{n}{2}}$ & $pr$ &  $p$ is prime, &   \\
                            & & & & & &   $r\mid (p^n-1)$  \\
                            & & & & & &   and $K=o(r)$\\
                            \hline

$17$ & $(\mu-1)N$ &  $N$ & $N$ & $N$ & $N$ & $1 < N \in \mathbb{N}$ is odd; & Thm. \ref{thm41}   \\
 &&&&&&       $\mu$ is the  smallest              \\
 &&&&&&       prime factor of $N$               \\
                                     \hline
$18$ & $(\mu-1)N$ &  $N-1$ & $N$ & $N$ & $N$ &  $1 < N \in \mathbb{N}$ is odd; & Thm. \ref{thm42}   \\
&&&&&&       $\mu$ is the  smallest              \\
&&&&&&       prime factor of $N$               \\
\bottomrule
\end{tabular}
\end{table}

After this introduction, Section \ref{S2} reviews basic definitions and results about characters and character sums over finite fields. We propose a general construction of QCSSs in Section \ref{S3}. Seven classes of asymptotically optimal periodic QCSSs are produced based on the proposed construction. Section \ref{S4} discusses our construction of two classes of asymptotically optimal periodic QCSSs with flexible parameters. Concluding remarks bring the paper to a close in Section \ref{S6}. All computations are done in {\tt MAGMA} V2.26-4 \cite{BJP97Magma}.

\section{Preliminaries}\label{S2}

Let $q=p^n$ be a prime power and let $\Ff_q$ stand for the finite field with $q$ elements. The trace mapping from $\Ff_q$ to $\Ff_p$ is given by
\[
\Tr_{q/p}(x)=x+x^p+\cdots+x^{p^{n-1}} \mbox{ for any } x\in\Ff_q.
\]
For each $a\in \mathbb{F}_q$, an {\it additive character} of $\mathbb{F}_q$ is defined by
\begin{equation}\label{eq:addchar}
\chi_a(x)=\xi_p^{\Tr_{q/p}(ax)} \mbox{, with }
\xi_p := e^{2\pi\sqrt{-1}/p}.
\end{equation}
The additive character $\chi_0(x)$ is the {\it trivial} one. The {\it conjugate} $\chi_a(x)^*$ of an additive character $\chi_a(x)$ of $\mathbb{F}_q$ is $\chi_a(-x)$. The additive characters of $\Ff_q$ form a group $\widehat{\Ff_q}$ under multiplication, which is defined by
\[
\chi_a(x) \, \chi_b(x)=\chi_{a+b}(x) \mbox{ for any } x\in\Ff_q.
\]
The groups $\widehat{\Ff_q}$ and $(\Ff_q,+)$ are isomorphic. The respective orthogonal relations of $(\Ff_q,+)$ and $\widehat{\Ff_q}$ are
\begin{equation}\label{eq:addort}
\sum_{g\in \Ff_q}\chi_a(g) =
\begin{cases}
0,& \mbox{if $a\neq 0$}\\
q,& \mbox{if $a=0$}
\end{cases}
\mbox{ and }
\sum_{a\in \Ff_q}\chi_a(g) =
\begin{cases}
0,& \mbox{if\ $g\neq 0$}\\
q,& \mbox{if\ $g=0$}
\end{cases}.
\end{equation}

Let $\Ff_q^*$ be the multiplicative group of $\Ff_q$. It is a cyclic group of order $q-1$. A generator of $\Ff_q^*$ is called a {\it primitive element} of $\Ff_q$. Let $\alpha$ be a primitive element of $\Ff_q$. For each $i \in[q-2]$, a {\it multiplicative character} $\varphi_i$ of $\Ff_q$ is defined by
\begin{equation}\label{eq:multchar}
\varphi_i(\alpha^j) = \xi_{q-1}^{i j} \mbox{, with }
j\in[q-2] \mbox{ and } \xi_{q-1} :=e^{2\pi\sqrt{-1}/(q-1)}.
\end{equation}
Multiplicative characters of $\mathbb{F}_q$ have similar properties as additive characters of $\Ff_q$. The orthogonal relations of multiplicative characters are given by
\begin{equation}\label{eq:multort}
\sum_{j=0}^{q-2}\varphi_i(\alpha^j) =
\begin{cases}
0,& \mbox{if $i\neq 0$}\\
q-1,& \mbox{if $i=0$}
\end{cases}
\mbox{ and }
\sum_{i=0}^{q-2} \varphi_i(\alpha^j) =
\begin{cases}
0,& \mbox{if\ $j\neq 0$}\\
q-1,& \mbox{if\ $j=0$}
\end{cases}.
\end{equation}

Let $\chi_a$ be an additive character and let $\varphi$ be a multiplicative character of $\Ff_q$. Their {\it Gauss sum} over $\Ff_q$ is
\begin{equation}\label{eq:Gauss}
G(\varphi,\chi_a) = \sum_{x\in \mathbb{F}_q^*} \varphi(x) \, {\chi_a}(x).
\end{equation}
It is immediate to infer that $|G(\varphi,\chi_a)| \leq q-1$. The following result gives the exact value depending on whether or not either character is trivial.
\begin{lem}{\rm \cite[Theorem 5.11]{field}}\label{gauss}
Given an additive character $\chi_a$ and a multiplicative character $\varphi$ of $\Ff_q$,
\begin{align*}
G(\varphi,\chi_a) &=
\begin{cases}
q-1, & \mbox{if } \varphi=\varphi_0 \mbox{ and } \chi_a=\chi_0, \\
-1,& \mbox{if } \varphi=\varphi_0 \mbox{ and }\chi_a\neq\chi_0,\\
0,& \mbox{if } \varphi\neq\varphi_0 \mbox{ and } \chi_a= \chi_0,
\end{cases}\\
\left|G(\varphi,\chi_a)\right| &= \sqrt{q}, \mbox{ if }
\varphi \neq \varphi_0 \mbox{ and } \chi_a \neq \chi_0.
\end{align*}
\end{lem}

\section{Asymptotically Optimal Periodic QCSSs from Periodic Sequence Sets}\label{S3}

We now propose a general construction of periodic QCSSs from interleaved sequences. The main insight is simple: \emph{A sequence set that asymptotically attains the Welch bound generates an asymptotically optimal periodic QCSS.}

\subsection{A general construction of periodic QCSSs}

The concept of {\it interleaved sequence} was introduced by Gong in \cite{Gong}. Given integers $u>1$ and $v>1$, the components of a sequence $\mathbf{s}=(s(0),s(1),\ldots,s(uv-1))$ of period $uv$ can be arranged into the $u\times v$ matrix
\[
\begin{pmatrix}
s_i(0) & s_i(1) & \cdots & s_i(v-1)\\
s_i(v) & s_i(v+1) & \cdots & s_i(v+v-1)\\
\vdots&\vdots &\ddots &\vdots\\
s_i((u-1)v) & s_i((u-1)v+1) & \cdots & s_i((u-1)v+v-1)
\end{pmatrix}.
\]
If each column of the matrix is either the zero sequence or a phase shift of a sequence with period $u$, then $\mathbf{s}$ is a $(u,v)$ interleaved sequence.

Based on Gong's interleaving technique, we arrange the components of a sequence of period $NK$ into a $K \times N$ matrix. Our general construction is given as Construction \ref{con1}.

\begin{algorithm}[H]
\caption{A general construction of periodic QCSSs from periodic sequence sets.}
\label{con1}
\begin{algorithmic}
\STATE{\textbf{Step 1.} Choose a periodic sequence set $\mathcal{S}=\{\mathbf{s}_i:i\in[M]\}$, with parameters $(n,M,\theta_{\rm max})$, that satisfies the following properties:
\begin{enumerate}
    \item The period is $n=NK$ for some integers $N>1$ and $K>1$.
    \item The family size of $\mathcal{S}$ is $M > K$.
\end{enumerate}
}
\STATE{\textbf{Step 2.} Rewrite each sequence $\mathbf{s}_i=(s_i(0),\cdots,s_i(n-1))$ of $\mathcal{S}$ as a $K\times N$ matrix of the form
\[
\cc(\mathbf{s}_i) =
\begin{pmatrix}
s_i(0) & s_i(0+K) & \cdots & s_i(0+K(N-1))\\
s_i(1) & s_i(1+K) & \cdots & s_i(1+K(N-1))\\
\vdots&\vdots &\ddots &\vdots\\
s_i(K-1) & s_i(K-1+K) & \cdots & s_i(K-1+K(N-1))
\end{pmatrix}.
\]
}
\STATE{\textbf{Step 3.} Generate the set $\cc(\mathcal{S}) := \left\{\cc(\mathbf{s}_0),\ldots,\cc(\mathbf{s}_{M-1})\right\}$.}
\end{algorithmic}
\end{algorithm}

Our next task is to determine the parameters of the output $\cc(\mathcal{S})$ of Construction \ref{con1}.

\begin{thm}\label{gcon}
The output set $\cc(\mathcal{S})$ of Construction \ref{con1} is a periodic $(M,K,N,\vartheta_{\rm max})$-QCSS with $\vartheta_{\rm max}\leq \theta_{\rm max}$.
\end{thm}
\begin{IEEEproof}
The steps in Construction \ref{con1} ensure that $\cc(\mathcal{S})$ is periodic, with length $N$, flock size $K$, and family size $M$. What remains is to bound its maximum periodic correlation magnitude. Let $\tau\in[N]$. For any two complementary sequences $\cc(\mathbf{s}_i)$ and $\cc(\mathbf{s}_j)$ of $\cc(\mathcal{S})$, we have
\begin{equation}
R_{\cc(\mathbf{s}_i),\cc(\mathbf{s}_j)}(\tau)
= \sum_{k=0}^{K-1}\sum_{t=0}^{N-1}s_i(k+Kt)s_j^*\left(k+K(t+\tau)\right)
= \sum_{x=0}^{n-1}s_i(x)s_j^*\left(x+K\tau\right) = R_{\mathbf{s}_i,\mathbf{s}_j}(K\tau).
\end{equation}
We conclude that $\vartheta_{\rm max}\leq \theta_{\rm max}$ by comparing
\begin{align*}
\vartheta_{\rm max} &=
\max_{i,j\in[M]} \left\{\left|R_{\cc(\mathbf{s}_i), \cc(\mathbf{s}_j)}(\tau_1)\right| \mbox{ with } \tau_1\in[N] \mbox{ when } i\neq j \mbox{ and }
\tau_1\in[N]\setminus\{0\} \mbox{ when } i=j\right\}
\mbox{ and}\\
\theta_{\rm max} &= \max_{i,j\in[M]} \left\{\left|R_{\mathbf{s}_i,\mathbf{s}_j}(\tau_2)\right| \mbox{ with } \tau_2\in[n] \mbox{ when } i \neq j \mbox{ and } \tau_2\in[n]\setminus\{0\} \mbox{ when } i=j\right\}.
\end{align*}
\end{IEEEproof}

To get a QCSS with low maximum periodic correlation magnitude, Theorem \ref{gcon} tells us to select a sequence set whose maximum periodic correlation magnitude is as low as possible. A sequence set $\mathcal{S}$ with parameters $(n,M,\theta_{\rm max})$ has a {\it low correlation} if $\theta_{\rm max} \leq \mu \sqrt{n}$ for a small positive constant $\mu$. If $\theta_{\rm max}$ approaches the Welch bound $n\sqrt{\frac{M-1}{nM-1}}$ as $M$ grows, then $\mathcal{S}$ is {\it asymptotically optimal with respect to the Welch bound} or, simply, {\it asymptotically optimal}. Many researchers have tried to construct sequence sets with low correlation. To our best knowledge, however, there are but a few such constructions. Those already available in the literature are listed in Table \ref{table2}.

\begin{table}[htb!]
\caption{The parameters of known asymptotically optimal sequence sets of period $n=NK$, family size $M$, and alphabet size $|\mathcal{A}|$.}
\label{table2}
\renewcommand{\arraystretch}{1.2}
\centering
\begin{tabular}{cccccll}
\toprule
No. & $n=NK$ & $M$ & $\theta_{\rm max}$ & $|\mathcal{A}|$ & Constraints            & References \\ \midrule
$1$ & $p$    & $p-2$       & $2+\sqrt{p}$   & $p-1$         & $p$ is odd prime    & \cite{Schmidt,Scholtz}  \\% \hline

$2$ & $p$    & $p$         & $\sqrt{p}$     & $p$           & $p \geq 5$ is prime   & \cite{Alltop}  \\% \hline
$3$ & $p^2$  & $p-1$       & $p$            & $p$           & $p$ is odd prime    & \cite{Frank}  \\% \hline

$4$ & $N$    & $\mu_{\min}$  & $\sqrt{N}$     & $N$           & $N=s\ell^2$ is odd and $\mu_{\min}$ is & \cite{Popovic}  \\% \hline
 &  &  & &   &  the smallest prime factor of $N$  &  \\% \hline

$5$ & $p^n-1$  & $p^n$     & $p^{\frac{n}{2}}+1$            & $p$           & $p$ is odd prime    & \cite{Sidelnikov}  \\% \hline
$6$ & $p^n-1$  & $p^{\frac{n}{2}}+1$     & $p^{\frac{n}{2}}+1$            & $p$           & $p=2$    & \cite{Kasami}  \\% \hline
$7$ & $p^n-1$  & $p^{\frac{n}{2}}$     & $p^{\frac{n}{2}}+1$            & $p$           & $p$ is odd prime    & \cite{Jang,Kumar,Liu,Moriuchi}  \\% \hline
$8$ & $p^n-1$  & $p^n+1$     & $p^{\frac{n}{2}}+1$            & $4$           & $p=2$     & \cite{Boztas,Tang}  \\
$9$ & $p(p^n-1)$  & $p^n$     & $p^{\frac{n+1}{2}}+p$         & $4$           & $p=2$     & \cite{Tang,Parampalli}  \\
$10$ & $p(p-1)$  & $p$     & $p$ & $p$  & $p$ is odd prime    & \cite{Chung}  \\
$11$ & $p^n-1$  & $r$     & $p^{\frac{n}{2}}$            & $pr$           & $p$ is prime and $r \mid (p^n-1)$    & \cite{D2}  \\% \hline
\bottomrule
\end{tabular}
\end{table}

Looking up Table \ref{table2}, we notice that the period and family size of an asymptotically optimal sequence set are functions of a single variable. Hence, we can write the parameters of such a sequence set as
\begin{equation}\label{eq:varx}
(n(x),M(x),\theta_{\rm max}) \mbox{, with }
\lim_{x \to +\infty} \frac{\theta_{\rm max}} {\sqrt{n(x)} \, \sqrt{\frac{(M(x)-1)n(x)}{n(x)M(x)-1}}}=1.
\end{equation}
We can then construct a QCSS that is asymptotically optimal by using an asymptotically optimal sequence set as the chosen $\mathcal{S}$ in {\bf Step 1} of Construction \ref{con1}.

\begin{thm}\label{thm3.2}
Let $\mathcal{S}$ be a sequence set with parameters $\left(n(x) : =N(x)K(x),M(x),\theta_{\rm max}\right)$ taken from Table \ref{table2}. Let $\cc(\mathcal{S})$ be a periodic $(M(x),K(x),N(x),\vartheta_{\rm max})$-QCSS built by Construction \ref{con1}. If
\[
\lim_{x\to +\infty} \frac{K(x)}{M(x)}=0,
\]
then $\cc(\mathcal{S})$ is asymptotically optimal.
\end{thm}
\begin{IEEEproof}
We start by observing that
\[
\theta_{\rm max} = \sqrt{n(x)}+c \mbox{, where } c=o\left(\sqrt{n(x)}\right).
\]
By Theorem \ref{gcon}, we have $\vartheta_{\rm max}\leq \sqrt{n(x)}+c$. We use the lower bound for $\vartheta_{\rm max}$ in (\ref{bound}) in combination with the given
assumption $\displaystyle{\lim_{x \to +\infty}\frac{K(x)}{M(x)}=0}$ and $n(x)=N(x) K(x)$, with $K(x)>1$, to derive
\begin{align*}
\lim_{x \to +\infty} \frac{\vartheta_{\rm max}}{K(x)N(x) \sqrt{\frac{\frac{M(x)}{K(x)}-1}{M(x)N(x)-1}}} &= \lim_{x \to +\infty} \frac{\vartheta_{\rm max}}{\sqrt{n(x)} \sqrt{\frac{M(x)N(x)-K(x)N(x)}{M(x)N(x)-1}}}\\
&\leq \lim_{x \to +\infty} \frac{\sqrt{n(x)}+c}{\sqrt{n(x)}} \sqrt{\frac{M(x)N(x)-1}{M(x)N(x)-K(x)N(x)}}\\
&= \lim_{x \to +\infty} \left(1+\frac{c}{\sqrt{n(x)}}\right) \sqrt{\frac{1-\frac{1}{M(x)N(x)}}{1-\frac{K(x)}{M(x)}}} = 1.
\end{align*}
The periodic QCSS $\cc(\mathcal{S})$ is asymptotically optimal since
\[
\frac{\vartheta_{\rm max}}{K(x)N(x) \sqrt{\frac{\frac{M(x)}{K(x)}-1}{M(x)N(x)-1}}} \geq 1 \implies
\lim_{x \to +\infty} \frac{\vartheta_{\rm max}}{K(x)N(x) \sqrt{\frac{\frac{M(x)}{K(x)}-1}{M(x)N(x)-1}}}=1.
\]
\end{IEEEproof}

\begin{remark}\label{remark1}
Theorem \ref{thm3.2} requires an asymptotically optimal sequence set whose length is the product of two positive integers $N>1$ and $K>1$, where $K$ is an infinitesimal of higher order than the family size $M$, that is, $K=o(M)$. This requirement rules out the first four classes of sequence sets in Entries $1$ to $4$ of Table \ref{table2}. All other entries in Table \ref{table2} can be deployed to construct asymptotically optimal QCSSs. Entries $10$ to $16$ in Table \ref{tabl1} are the seven classes of asymptotically optimal QCSSs generated by applying Theorem \ref{thm3.2} on Entries $5$ to $11$ in Table \ref{table2}, in that order.
\end{remark}

\subsection{A specific construction of asymptotically optimal periodic QCSSs}

We showcase the power of Theorem \ref{thm3.2} in a specific construction of asymptotically optimal periodic QCSSs. Zhou, Helleseth, and Parampalli in \cite{D2} combined additive and multiplicative characters to propose a family of asymptotically optimal sequences with respect to the Welch bound. This family is Entry $11$ in Table \ref{table2}.

\begin{prop}{\rm \cite{D2}}\label{prop1}
Let $\chi_1$ be the canonical additive character and let $\varphi_j$ be a multiplicative character of $\Ff_q$ defined as in (\ref{eq:multchar}). For a given primitive element $\alpha$ of $\Ff_q$, let
\[
\mathbf{s}_j :=\left(\chi_1(\alpha^t) \, \varphi_j(\alpha^t)\right)_{t=0}^{q-2} \mbox{ for each } j\in[q-2].
\]
The sequence set $\mathcal{S}_1 = \{\mathbf{s}_j : j\in[q-2]\}$, with parameters $(q-1,q-1,\sqrt{q})$, asymptotically meets the Welch bound.
\end{prop}

Applying Proposition \ref{prop1} to Theorems \ref{gcon} and \ref{thm3.2} yields the next result.

\begin{thm}\label{thm3.3}
Let $\mathcal{S}_1$ be the sequence set defined in Proposition \ref{prop1}. If $q-1=NK$, with $1<K<q-1$, then the set $\cc(\mathcal{S}_1)$ constructed by Theorem \ref{gcon} is a $(q-1,K,N,\sqrt{q})$-QCSS. Moreover, if $\lim_{q\to +\infty}\frac{K}{q-1}=0$, then $\cc(\mathcal{S}_1)$ asymptotically meets the bound in (\ref{bound}).
\end{thm}
\begin{IEEEproof}
Since $q-1>K$, Theorem \ref{gcon} implies that $\cc(\mathcal{S}_1)$ is a QCSS with set size $q-1$, flock size $K$ and sequence length $N$. We present the computation of $|R_{\cc(\mathbf{s}_i),\cc(\mathbf{s}_j)}(\tau)|$ of $\cc(\mathcal{S}_1)$ in two cases.

\begin{enumerate}[wide, itemsep=0pt, leftmargin =0pt,widest={{\bf Case $2$}}]
\item[{\bf Case $1$}:] If $i=j$ and $\tau\in[N]\setminus\{0\}$, then
\begin{align*}
R_{\cc(\mathbf{s}_i),\cc(\mathbf{s}_i)}(\tau)
&= \sum_{k=0}^{K-1} \sum_{t=0}^{N-1} \chi_1\left(\alpha^{k+Kt}\right) \varphi_i\left(\alpha^{k+Kt}\right) \chi_1\left(-\alpha^{k+K(t+\tau)}\right) \varphi_i\left(\alpha^{-k-K(t+\tau)}\right)\\
&= \varphi_i\left(\alpha^{-K\tau}\right) \sum_{k=0}^{K-1} \sum_{t=0}^{N-1}
\chi_1\left(\alpha^{k+Kt}(1-\alpha^{K\tau})\right)
= \varphi_i\left(\alpha^{-K\tau}\right) \sum_{x=0}^{q-2} \chi_1\left(\alpha^x(1-\alpha^{K\tau})\right).
\end{align*}
Due to $\tau\in[N]\setminus\{0\}$, we obtain $\alpha^{K\tau} \neq 1$. By the orthogonal relations in (\ref{eq:addort}),
\[
\left|R_{\cc(\mathbf{s}_i),\cc(\mathbf{s}_i)}(\tau)\right|=\left|\sum_{z\in\Ff_q}\chi_{(1-\alpha^{K\tau})}\left(z\right)-1\right|=1.
\]
\item[{\bf Case $2$}:]If $i\neq j$ and $\tau\in[N]$, then
\begin{align*}
R_{\cc(\mathbf{s}_i),\cc(\mathbf{s}_j)}(\tau)
&= \sum_{k=0}^{K-1}\sum_{t=0}^{N-1} \chi_1\left(\alpha^{k+Kt}\right) \varphi_i\left(\alpha^{k+Kt}\right) \chi_1\left(-\alpha^{k+K(t+\tau)}\right) \varphi_j\left(\alpha^{-k-K(t+\tau)}\right)\\
&= \varphi_j\left(\alpha^{-K\tau}\right) \sum_{k=0}^{K-1} \sum_{t=0}^{N-1} \chi_1\left(\alpha^{k+Kt}(1-\alpha^{K\tau})\right) \varphi_i\left(\alpha^{k+Kt}\right) \varphi_{-j}\left(\alpha^{k+Kt}\right)\\
&= \varphi_j\left(\alpha^{-K\tau}\right) \sum_{x=0}^{q-2} \chi_1\left(\alpha^x(1-\alpha^{K\tau})\right) \varphi_{i-j}\left(\alpha^{x}\right).
\end{align*}
When $\tau=0$, one uses the fact that $i\neq j$ and the orthogonal relations in (\ref{eq:multort}) to infer
\[
R_{\cc(\mathbf{s}_i),\cc(\mathbf{s}_j)}(\tau) =\sum_{x=0}^{q-2}\varphi_{i-j}\left(\alpha^{x}\right)=0.
\]
When $\tau \neq 0$, Lemma \ref{gauss} implies $\alpha^{K\tau}\neq 1$. Hence,
\begin{equation}
\left|R_{\cc(\mathbf{s}_i),\cc(\mathbf{s}_j)}(\tau)\right| = \left|\sum_{x=0}^{q-2}\chi_1\left(\alpha^x(1-\alpha^{K\tau})\right)\varphi_{i-j}\left(\alpha^{x}\right)\right|
= \left|\sum_{z\in\Ff_q^*}\chi_{\left(1-\alpha^{K\tau}\right)}(z)\varphi_{i-j}(z)\right| = \sqrt{q}.
\end{equation}
\end{enumerate}
Therefore, $\cc(\mathcal{S}_1)$ has $\vartheta_{\rm max} = \sqrt{q}$. The bound in (\ref{bound}) implies
\[
\vartheta_{\rm opt} = (q-1) \sqrt{\frac{\frac{q-1}{K}-1}{(q-1)N-1}} = (q-1)\sqrt{\frac{(q-1)-K}{(q-1)^2-K}}.
\]
The conclusion that $\cc(\mathcal{S}_1)$ is asymptotically optimal follows from
\[
\lim_{q\to +\infty}\frac{\sqrt{q}}{\vartheta_{\rm opt}}=\lim_{q\to +\infty}\sqrt{\frac{q(q-1)^2-qK}{(q-1)^3-(q-1)^2K}}=1.
\]
\end{IEEEproof}

The next example illustrates Theorem \ref{thm3.3}.
\begin{example}
Let $q=16$, $N=5$, and $K=3$. Theorem \ref{thm3.3} gives us  a periodic $(15,3,5,4)$-QCSS. It consists of $15$ matrices $\cc^0$ to $\cc^{14}$ below. We replace each original entry of $\cc^i$, for all $i \in [15]$, by its corresponding \emph{exponent} when expressed as a \emph{power} of $\xi_{30} :=e^{2\pi\sqrt{-1}/30}$. The first row of $\cc^{0}$, for instance, is originally
\[
\left(\xi_{30}^0, \xi_{30}^{15}, \xi_{30}^{15}, \xi_{30}^{15}, \xi_{30}^{15}\right) =
\left(1,e^{\pi\sqrt{-1}}, e^{\pi\sqrt{-1}},e^{\pi\sqrt{-1}},e^{\pi\sqrt{-1}}\right),
\]
but is succinctly presented as $(0,15,15,15,15)$.
\[
\cc^0=\begin{pmatrix}
 0& 15& 15& 15& 15\\
 0& 0& 15& 0& 15 \\
 0& 0& 0& 15& 15
\end{pmatrix}, \quad
\cc^1=\begin{pmatrix}
  0& 21& 27& 3& 9  \\
  2& 8& 29& 20& 11  \\
  4& 10& 16& 7& 13
\end{pmatrix}, \quad
\cc^2=\begin{pmatrix}
  0& 27& 9& 21& 3  \\
  4& 16& 13& 10& 7  \\
  8& 20& 2& 29& 11
\end{pmatrix},
\]

\[
\cc^3=\begin{pmatrix}
  0& 3& 21& 9& 27  \\
  6& 24& 27& 0& 3  \\
  12& 0& 18& 21& 9
\end{pmatrix}, \quad
\cc^4=\begin{pmatrix}
  0& 9& 3& 27& 21  \\
  8& 2& 11& 20& 29  \\
  16& 10& 4& 13& 7
\end{pmatrix}, \quad
\cc^5=\begin{pmatrix}
  0& 15& 15& 15& 15  \\
  10& 10& 25& 10& 25  \\
  20& 20& 20& 5& 5
\end{pmatrix},
\]

\[
\cc^6=\begin{pmatrix}
  0& 21& 27& 3& 9  \\
  12& 18& 9& 0& 21  \\
  24& 0& 6& 27& 3
\end{pmatrix},\quad
\cc^7=\begin{pmatrix}
  0& 27& 9& 21& 3  \\
  14& 26& 23& 20& 17  \\
  28& 10& 22& 19& 1
\end{pmatrix}, \quad
\cc^8=\begin{pmatrix}
  0& 3& 21& 9& 27  \\
  16& 4& 7& 10& 13  \\
  2& 20& 8& 11& 29
\end{pmatrix},
\]

\[
\cc^9=\begin{pmatrix}
  0& 9& 3& 27& 21  \\
  18& 12& 21& 0& 9  \\
  6& 0& 24& 3& 27
\end{pmatrix}, \quad
\cc^{10}=\begin{pmatrix}
  0& 15& 15& 15& 15  \\
  20& 20& 5& 20& 5  \\
  10& 10& 10& 25& 25
\end{pmatrix}, \quad
\cc^{11}=\begin{pmatrix}
  0& 21& 27& 3& 9  \\
  22& 28& 19& 10& 1  \\
  14& 20& 26& 17& 23
\end{pmatrix},
\]

\[
\cc^{12}=\begin{pmatrix}
  0& 27& 9& 21& 3  \\
  24& 6& 3& 0& 27 \\
  18& 0& 12& 9& 21
\end{pmatrix}, \quad
\cc^{13}=\begin{pmatrix}
  0& 3& 21& 9& 27  \\
  26& 14& 17& 20& 23  \\
  22& 10& 28& 1& 19
\end{pmatrix}, \quad
\cc^{14}=\begin{pmatrix}
  0& 9& 3& 27& 21  \\
  28& 22& 1& 10& 19  \\
  26& 20& 14& 23& 17
\end{pmatrix}.
\]\QEDB
\end{example}

\section{Two New Families of Asymptotically Optimal Periodic QCSSs with Flexible Parameters}\label{S4}

Popovitch constructed the generalized chirp-like polyphase sequence sets, listed as Entry $4$ in Table \ref{table2}, in \cite{Popovic}, based on the so-called Zadoff-Chu polyphase sequences from \cite{Frank}. The parameters $(N,\mu_{\min},\sqrt{N})$ of the chirp-like sequence sets are very flexible. We have stated in Remark \ref{remark1} that using the generalized chirp-like polyphase sequence sets as ingredients in Theorem \ref{thm3.2} does not yield asymptotically optimal periodic QCSSs.

In this section, we present two new families of periodic asymptotically optimal QCSSs with flexible parameters. We start with a lemma that will play a key role in the computation of the periodic correlations.

\begin{lem}{\rm \cite[Chp. 8]{Silverman}}\label{lem41}
Let $a$, $b$, and $m>1$ be integers with $g:=\gcd(a,m)$. If $g \mid b$, then the congruence $ay \equiv b \pmod m$ has exactly $g$ incongruent solutions for $y$.
\end{lem}

Let $N>1$ be an odd integer and let $\mu_{\min}$ be the smallest prime factor of $N$. Let $\xi_N$ be a primitive $N^{\rm th}$ root of unity. Let $\rho$ be a bijection from $[N]$ to $[N]$. For each $a\in [\mu_{\min}]\setminus\{0\}$ and $b\in[N]$, we design the two-dimensional $N \times N$ matrix
\[
\mathcal{C}^{a,b} =
\begin{pmatrix}
1 & \xi_N^{a\rho(0)+b} & \cdots & \xi_N^{(a\rho(0)+b)(N-1)}\\
1 & \xi_N^{a\rho(1)+b} & \cdots & \xi_N^{(a\rho(1)+b)(N-1)}\\
\vdots&\vdots &\ddots &\vdots\\
1 & \xi_N^{a\rho(N-1)+b} & \cdots & \xi_N^{(a\rho(N-1)+b)(N-1)}\\
\end{pmatrix}.
\]
It gives us the complementary sequence set
\begin{equation}\label{con2}
\mathfrak{C}=\{\mathcal{C}^{a,b}:a \in [\mu_{\min}]\setminus\{0\} \mbox{ and } b\in[N]\}.
\end{equation}

\begin{thm}\label{thm41}
Let $N>1$ be an odd integer and let $\mu_{\min}$ be the smallest prime factor of $N$. Then the set $\mathfrak{C}$ in (\ref{con2}) is an asymptotically optimal QCSS with parameters $((\mu_{\min}-1)N,N,N,N)$.
\end{thm}
\begin{IEEEproof}
By how it is defined, $\mathfrak{C}$ has flock size $N$ and sequence length $N$. To determine  $\left|R_{\cc^{a_1,b_1},\cc^{a_2,b_2}}(\tau)\right|$ of $\mathfrak{C}$, we verify that
\begin{equation*}
R_{\cc^{a_1,b_1},\cc^{a_2,b_2}}(\tau)
= \sum_{i=0}^{N-1} \sum_{j=0}^{N-1} \xi_N^{(a_1\rho(i)+b_1)j-(a_2\rho(i)+b_2)(j+\tau)} = \xi_N^{-b_2\tau} \sum_{i=0}^{N-1}\xi_N^{-a_2\rho(i)\tau} \sum_{j=0}^{N-1}\xi_N^{\left((a_1-a_2) \rho(i)+b_1-b_2\right)j}.
\end{equation*}
We proceed by cases.

\begin{enumerate}[wide, itemsep=0pt, leftmargin =0pt,widest={{\bf Case $2$}}]

\item[{\bf Case $1$}:] If $a_1=a_2$, $b_1=b_2$, and $\tau\in[N]\setminus\{0\}$, then
\[
R_{\cc^{a_1,b_1},\cc^{a_2,b_2}}(\tau)=N\xi_N^{-b_2\tau}\sum_{i=0}^{N-1}\xi_N^{-a_2\rho(i)\tau}.
\]
Since $\tau\in[N]\setminus\{0\}$ and $a_2 \in [\mu_{\min}]\setminus\{0\}$, we know that  $a_2\tau \not\equiv 0 \pmod N$. Hence, $R_{\cc^{a_1,b_1},\cc^{a_2,b_2}}(\tau)=0$.

\item[{\bf Case $2$}:]If $a_1=a_2$, $b_1\neq b_2$, and $\tau\in[N]$, then
\[
R_{\cc^{a_1,b_1},\cc^{a_2,b_2}}(\tau)=\xi_N^{-b_2\tau}\sum_{i=0}^{N-1}\xi_N^{-a_2\rho(i)\tau}\sum_{j=0}^{N-1}\xi_N^{\left(b_1-b_2\right)j}=0.
\]

\item[{\bf Case $3$}:]Assume that $\rho(i_0)=0$. If $a_1\neq a_2$, $b_1=b_2$, and $\tau\in[N]$, then
\begin{align*}
R_{\cc^{a_1,b_1},\cc^{a_2,b_2}}(\tau) &= \xi_N^{-b_2\tau} \sum_{i=0}^{N-1}\xi_N^{-a_2\rho(i)\tau} \sum_{j=0}^{N-1}\xi_N^{\left((a_1-a_2)\rho(i)\right)j}\\
&= \xi_N^{-b_2\tau}\xi_N^{-a_2\rho(i_0)\tau} \sum_{j=0}^{N-1}\xi_N^{\left((a_1-a_2)\rho(i_0)\right)j}+ \xi_N^{-b_2\tau}\sum_{i=0,i\neq i_0}^{N-1} \xi_N^{-a_2\rho(i)\tau} \sum_{j=0}^{N-1}\xi_N^{\left((a_1-a_2)\rho(i)\right)j}\\
&= N\xi_N^{-b_2\tau}.
\end{align*}
Thus, $\left|R_{\cc^{a_1,b_1},\cc^{a_2,b_2}}(\tau)\right|=N$.
\item[{\bf Case $4$}:]If $a_1\neq a_2$, $b_1\neq b_2$, and $\tau\in[N]$, then
\[
R_{\cc^{a_1,b_1},\cc^{a_2,b_2}}(\tau)=\xi_N^{-b_2\tau}\sum_{i=0}^{N-1}\xi_N^{-a_2\rho(i)\tau}\sum_{j=0}^{N-1}\xi_N^{\left((a_1-a_2)\rho(i)+b_1-b_2\right)j}.
\]
By Lemma \ref{lem41}, there exists a unique solution $\ell$ to the congruence $(a_1-a_2)\rho(i)\equiv b_2-b_1 \pmod N$, since $\gcd(a_1-a_2,N)=1$. Hence,
\begin{align*}
\left|R_{\cc^{a_1,b_1},\cc^{a_2,b_2}}(\tau)\right|
&= \left|\xi_N^{-b_2\tau} \xi_N^{-a_2\rho(\ell)\tau} \sum_{j=0}^{N-1}1+\xi_N^{-b_2\tau} \sum_{i=0,i\neq \ell}^{N-1}\xi_N^{-a_2\rho(i)\tau} \sum_{j=0}^{N-1}\xi_N^{\left((a_1-a_2) \rho(i)+b_1-b_2\right)j}\right|\\
&= \left|N\xi_N^{-b_2\tau}\xi_N^{-a_2\rho(\ell)\tau}\right| = N.
\end{align*}
\end{enumerate}
Since the above four cases cover all possibilities, the maximum periodic correlation magnitude is $\vartheta_{\rm max}=N$. It is clear that  $\left|R_{\cc^{a_1,b_1},\cc^{a_2,b_2}}(\tau)\right|=N^2$ if and only if $a_1=a_2$, $b_1=b_2$, and $\tau=0$. Hence, the family size of $\mathfrak{C}$ is $M=(\mu_{\min}-1)N$. By the bound in (\ref{bound}),
\[
\vartheta_{\rm opt}=N^2\sqrt{\frac{\mu_{\min}-2}{(\mu_{\min}-1)N^2-1}}.
\]
The conclusion that $\mathfrak{C}$ is asymptotically optimal follows once we confirm that
\[
\lim_{\mu_{\min} \to +\infty} \frac{N}{\vartheta_{\rm opt}}
= \lim_{\mu_{\min} \to +\infty} \sqrt{\frac{(\mu_{\min}-1)N^2-1}{(\mu_{\min}-2)N^2}} =1.
%= \lim_{\mu_{\min} \to +\infty} \sqrt{\frac{1-\frac{1}{(\mu_{\min}-1)N^2}} {\frac{1-2/\mu_{\min}}{1-1/\mu_{\min}}}} = 1.
\]
\end{IEEEproof}

\begin{example}
Let $N=9$ and $\rho(x)=x$. The $18$ complementary sequences expressed as matrices $\cc^0$ to $\cc^{17}$ below form the periodic QCSS $\mathfrak{C}$ constructed by Theorem \ref{thm41}, with $\vartheta_{\rm max} =9$. Each original entry of $\cc^i$ is replaced by the corresponding exponent when expressed as a power of $\xi_{9}:= e^{2\pi\sqrt{-1}/9}$.
\[
\cc^0=
\begin{pmatrix}
\begin{smallmatrix}
  0& 0& 0& 0& 0& 0& 0& 0& 0  \\
  0& 1& 2& 3& 4& 5& 6& 7& 8  \\
  0& 2& 4& 6& 8& 1& 3& 5& 7  \\
  0& 3& 6& 0& 3& 6& 0& 3& 6  \\
  0& 4& 8& 3& 7& 2& 6& 1& 5  \\
  0& 5& 1& 6& 2& 7& 3& 8& 4  \\
  0& 6& 3& 0& 6& 3& 0& 6& 3  \\
  0& 7& 5& 3& 1& 8& 6& 4& 2  \\
  0& 8& 7& 6& 5& 4& 3& 2& 1
\end{smallmatrix}
\end{pmatrix}, \quad
\cc^1=\begin{pmatrix}
\begin{smallmatrix}
  0& 0& 0& 0& 0& 0& 0& 0& 0  \\
  0& 2& 4& 6& 8& 1& 3& 5& 7  \\
  0& 4& 8& 3& 7& 2& 6& 1& 5  \\
  0& 6& 3& 0& 6& 3& 0& 6& 3  \\
  0& 8& 7& 6& 5& 4& 3& 2& 1  \\
  0& 1& 2& 3& 4& 5& 6& 7& 8  \\
  0& 3& 6& 0& 3& 6& 0& 3& 6  \\
  0& 5& 1& 6& 2& 7& 3& 8& 4  \\
  0& 7& 5& 3& 1& 8& 6& 4& 2
\end{smallmatrix}
\end{pmatrix}, \quad
\cc^2=\begin{pmatrix}
\begin{smallmatrix}
  0& 1& 2& 3& 4& 5& 6& 7& 8  \\
  0& 2& 4& 6& 8& 1& 3& 5& 7  \\
  0& 3& 6& 0& 3& 6& 0& 3& 6  \\
  0& 4& 8& 3& 7& 2& 6& 1& 5  \\
  0& 5& 1& 6& 2& 7& 3& 8& 4  \\
  0& 6& 3& 0& 6& 3& 0& 6& 3  \\
  0& 7& 5& 3& 1& 8& 6& 4& 2  \\
  0& 8& 7& 6& 5& 4& 3& 2& 1  \\
  0& 0& 0& 0& 0& 0& 0& 0& 0
\end{smallmatrix}
\end{pmatrix},
\]
\[
\cc^3=\begin{pmatrix}
\begin{smallmatrix}
  0& 1& 2& 3& 4& 5& 6& 7& 8  \\
  0& 3& 6& 0& 3& 6& 0& 3& 6  \\
  0& 5& 1& 6& 2& 7& 3& 8& 4  \\
  0& 7& 5& 3& 1& 8& 6& 4& 2  \\
  0& 0& 0& 0& 0& 0& 0& 0& 0  \\
  0& 2& 4& 6& 8& 1& 3& 5& 7  \\
  0& 4& 8& 3& 7& 2& 6& 1& 5  \\
  0& 6& 3& 0& 6& 3& 0& 6& 3  \\
  0& 8& 7& 6& 5& 4& 3& 2& 1
\end{smallmatrix}
\end{pmatrix}, \quad
\cc^4=\begin{pmatrix}
\begin{smallmatrix}
  0& 2& 4& 6& 8& 1& 3& 5& 7  \\
  0& 3& 6& 0& 3& 6& 0& 3& 6  \\
  0& 4& 8& 3& 7& 2& 6& 1& 5  \\
  0& 5& 1& 6& 2& 7& 3& 8& 4  \\
  0& 6& 3& 0& 6& 3& 0& 6& 3  \\
  0& 7& 5& 3& 1& 8& 6& 4& 2  \\
  0& 8& 7& 6& 5& 4& 3& 2& 1  \\
  0& 0& 0& 0& 0& 0& 0& 0& 0  \\
  0& 1& 2& 3& 4& 5& 6& 7& 8
\end{smallmatrix}
\end{pmatrix}, \quad
\cc^5=\begin{pmatrix}
\begin{smallmatrix}
  0& 2& 4& 6& 8& 1& 3& 5& 7  \\
  0& 4& 8& 3& 7& 2& 6& 1& 5  \\
  0& 6& 3& 0& 6& 3& 0& 6& 3  \\
  0& 8& 7& 6& 5& 4& 3& 2& 1  \\
  0& 1& 2& 3& 4& 5& 6& 7& 8  \\
  0& 3& 6& 0& 3& 6& 0& 3& 6  \\
  0& 5& 1& 6& 2& 7& 3& 8& 4  \\
  0& 7& 5& 3& 1& 8& 6& 4& 2  \\
  0& 0& 0& 0& 0& 0& 0& 0& 0
\end{smallmatrix}
\end{pmatrix},
\]
\[
\cc^6=\begin{pmatrix}
\begin{smallmatrix}
  0& 3& 6& 0& 3& 6& 0& 3& 6  \\
  0& 4& 8& 3& 7& 2& 6& 1& 5  \\
  0& 5& 1& 6& 2& 7& 3& 8& 4  \\
  0& 6& 3& 0& 6& 3& 0& 6& 3  \\
  0& 7& 5& 3& 1& 8& 6& 4& 2  \\
  0& 8& 7& 6& 5& 4& 3& 2& 1  \\
  0& 0& 0& 0& 0& 0& 0& 0& 0  \\
  0& 1& 2& 3& 4& 5& 6& 7& 8  \\
  0& 2& 4& 6& 8& 1& 3& 5& 7
\end{smallmatrix}
\end{pmatrix}, \quad
\cc^7=\begin{pmatrix}
\begin{smallmatrix}
  0& 3& 6& 0& 3& 6& 0& 3& 6  \\
  0& 5& 1& 6& 2& 7& 3& 8& 4  \\
  0& 7& 5& 3& 1& 8& 6& 4& 2  \\
  0& 0& 0& 0& 0& 0& 0& 0& 0  \\
  0& 2& 4& 6& 8& 1& 3& 5& 7  \\
  0& 4& 8& 3& 7& 2& 6& 1& 5  \\
  0& 6& 3& 0& 6& 3& 0& 6& 3  \\
  0& 8& 7& 6& 5& 4& 3& 2& 1  \\
  0& 1& 2& 3& 4& 5& 6& 7& 8
\end{smallmatrix}
\end{pmatrix}, \quad
\cc^8=\begin{pmatrix}
\begin{smallmatrix}
  0& 4& 8& 3& 7& 2& 6& 1& 5  \\
  0& 5& 1& 6& 2& 7& 3& 8& 4  \\
  0& 6& 3& 0& 6& 3& 0& 6& 3  \\
  0& 7& 5& 3& 1& 8& 6& 4& 2  \\
  0& 8& 7& 6& 5& 4& 3& 2& 1  \\
  0& 0& 0& 0& 0& 0& 0& 0& 0  \\
  0& 1& 2& 3& 4& 5& 6& 7& 8  \\
  0& 2& 4& 6& 8& 1& 3& 5& 7  \\
  0& 3& 6& 0& 3& 6& 0& 3& 6
\end{smallmatrix}
\end{pmatrix},
\]
\[
\cc^9=\begin{pmatrix}
\begin{smallmatrix}
  0& 4& 8& 3& 7& 2& 6& 1& 5  \\
  0& 6& 3& 0& 6& 3& 0& 6& 3  \\
  0& 8& 7& 6& 5& 4& 3& 2& 1  \\
  0& 1& 2& 3& 4& 5& 6& 7& 8  \\
  0& 3& 6& 0& 3& 6& 0& 3& 6  \\
  0& 5& 1& 6& 2& 7& 3& 8& 4  \\
  0& 7& 5& 3& 1& 8& 6& 4& 2  \\
  0& 0& 0& 0& 0& 0& 0& 0& 0  \\
  0& 2& 4& 6& 8& 1& 3& 5& 7
\end{smallmatrix}
\end{pmatrix}, \quad
\cc^{10}=\begin{pmatrix}
\begin{smallmatrix}
  0& 5& 1& 6& 2& 7& 3& 8& 4  \\
  0& 6& 3& 0& 6& 3& 0& 6& 3  \\
  0& 7& 5& 3& 1& 8& 6& 4& 2  \\
  0& 8& 7& 6& 5& 4& 3& 2& 1  \\
  0& 0& 0& 0& 0& 0& 0& 0& 0  \\
  0& 1& 2& 3& 4& 5& 6& 7& 8  \\
  0& 2& 4& 6& 8& 1& 3& 5& 7  \\
  0& 3& 6& 0& 3& 6& 0& 3& 6  \\
  0& 4& 8& 3& 7& 2& 6& 1& 5
\end{smallmatrix}
\end{pmatrix}, \quad
\cc^{11}=\begin{pmatrix}
\begin{smallmatrix}
  0& 5& 1& 6& 2& 7& 3& 8& 4  \\
  0& 7& 5& 3& 1& 8& 6& 4& 2  \\
  0& 0& 0& 0& 0& 0& 0& 0& 0  \\
  0& 2& 4& 6& 8& 1& 3& 5& 7  \\
  0& 4& 8& 3& 7& 2& 6& 1& 5  \\
  0& 6& 3& 0& 6& 3& 0& 6& 3  \\
  0& 8& 7& 6& 5& 4& 3& 2& 1  \\
  0& 1& 2& 3& 4& 5& 6& 7& 8  \\
  0& 3& 6& 0& 3& 6& 0& 3& 6
\end{smallmatrix}
\end{pmatrix},
\]
\[
\cc^{12}=\begin{pmatrix}
\begin{smallmatrix}
  0& 6& 3& 0& 6& 3& 0& 6& 3  \\
  0& 7& 5& 3& 1& 8& 6& 4& 2  \\
  0& 8& 7& 6& 5& 4& 3& 2& 1  \\
  0& 0& 0& 0& 0& 0& 0& 0& 0  \\
  0& 1& 2& 3& 4& 5& 6& 7& 8  \\
  0& 2& 4& 6& 8& 1& 3& 5& 7  \\
  0& 3& 6& 0& 3& 6& 0& 3& 6  \\
  0& 4& 8& 3& 7& 2& 6& 1& 5  \\
  0& 5& 1& 6& 2& 7& 3& 8& 4
\end{smallmatrix}
\end{pmatrix}, \quad
\cc^{13}=\begin{pmatrix}
\begin{smallmatrix}
  0& 6& 3& 0& 6& 3& 0& 6& 3  \\
  0& 8& 7& 6& 5& 4& 3& 2& 1  \\
  0& 1& 2& 3& 4& 5& 6& 7& 8  \\
  0& 3& 6& 0& 3& 6& 0& 3& 6  \\
  0& 5& 1& 6& 2& 7& 3& 8& 4  \\
  0& 7& 5& 3& 1& 8& 6& 4& 2  \\
  0& 0& 0& 0& 0& 0& 0& 0& 0  \\
  0& 2& 4& 6& 8& 1& 3& 5& 7  \\
  0& 4& 8& 3& 7& 2& 6& 1& 5
\end{smallmatrix}
\end{pmatrix}, \quad
\cc^{14}=\begin{pmatrix}
\begin{smallmatrix}
  0& 7& 5& 3& 1& 8& 6& 4& 2  \\
  0& 8& 7& 6& 5& 4& 3& 2& 1  \\
  0& 0& 0& 0& 0& 0& 0& 0& 0  \\
  0& 1& 2& 3& 4& 5& 6& 7& 8  \\
  0& 2& 4& 6& 8& 1& 3& 5& 7  \\
  0& 3& 6& 0& 3& 6& 0& 3& 6  \\
  0& 4& 8& 3& 7& 2& 6& 1& 5  \\
  0& 5& 1& 6& 2& 7& 3& 8& 4  \\
  0& 6& 3& 0& 6& 3& 0& 6& 3
\end{smallmatrix}
\end{pmatrix},
\]
\[
\cc^{15}=\begin{pmatrix}
\begin{smallmatrix}
  0& 7& 5& 3& 1& 8& 6& 4& 2  \\
  0& 0& 0& 0& 0& 0& 0& 0& 0  \\
  0& 2& 4& 6& 8& 1& 3& 5& 7  \\
  0& 4& 8& 3& 7& 2& 6& 1& 5  \\
  0& 6& 3& 0& 6& 3& 0& 6& 3  \\
  0& 8& 7& 6& 5& 4& 3& 2& 1  \\
  0& 1& 2& 3& 4& 5& 6& 7& 8  \\
  0& 3& 6& 0& 3& 6& 0& 3& 6  \\
  0& 5& 1& 6& 2& 7& 3& 8& 4
\end{smallmatrix}
\end{pmatrix}, \quad
\cc^{16}=\begin{pmatrix}
\begin{smallmatrix}
  0& 8& 7& 6& 5& 4& 3& 2& 1  \\
  0& 0& 0& 0& 0& 0& 0& 0& 0  \\
  0& 1& 2& 3& 4& 5& 6& 7& 8  \\
  0& 2& 4& 6& 8& 1& 3& 5& 7  \\
  0& 3& 6& 0& 3& 6& 0& 3& 6  \\
  0& 4& 8& 3& 7& 2& 6& 1& 5  \\
  0& 5& 1& 6& 2& 7& 3& 8& 4  \\
  0& 6& 3& 0& 6& 3& 0& 6& 3  \\
  0& 7& 5& 3& 1& 8& 6& 4& 2
\end{smallmatrix}
\end{pmatrix}, \quad
\cc^{17}=\begin{pmatrix}
\begin{smallmatrix}
  0& 8& 7& 6& 5& 4& 3& 2& 1  \\
  0& 1& 2& 3& 4& 5& 6& 7& 8  \\
  0& 3& 6& 0& 3& 6& 0& 3& 6  \\
  0& 5& 1& 6& 2& 7& 3& 8& 4  \\
  0& 7& 5& 3& 1& 8& 6& 4& 2  \\
  0& 0& 0& 0& 0& 0& 0& 0& 0  \\
  0& 2& 4& 6& 8& 1& 3& 5& 7  \\
  0& 4& 8& 3& 7& 2& 6& 1& 5  \\
  0& 6& 3& 0& 6& 3& 0& 6& 3
\end{smallmatrix}
\end{pmatrix}.
\]
\QEDB
\end{example}

Deleting the first row of each element of $\mathfrak{C}$ from (\ref{con2}) gives us the second construction of asymptotically optimal periodic QCSSs. Let $N>1$ be an odd integer and let $\mu_{\min}$ be the smallest prime factor of $N$. Let $\rho$ be a permutation on $[N]$ that fixes $0$. Define the complementary sequence set
\begin{equation}\label{con3}
\widehat{\mathfrak{C}} := \left\{\widehat{\mathcal{C}}^{a,b} : a \in [\mu_{\min}]\setminus\{0\} \mbox{ and } b\in[N]\right\},
\end{equation}
where
\[
\widehat{\mathcal{C}}^{a,b}=
\begin{pmatrix}
1 & \xi_N^{a\rho(1)+b} & \cdots & \xi_N^{(a\rho(1)+b)(N-1)}\\
\vdots&\vdots &\ddots &\vdots\\
1 & \xi_N^{a\rho(N-1)+b} & \cdots & \xi_N^{(a\rho(N-1)+b)(N-1)}\\
\end{pmatrix}
\]
is a two-dimensional $(N-1)\times N$ matrix.

\begin{thm}\label{thm42}
Let $N>1$ be an odd integer and let $\mu_{\min}$ be the smallest prime factor of $N$. Let $\rho$ be a permutation on $[N]$ that fixes $0$. Then the set $\widehat{\mathfrak{C}}$ as defined in (\ref{con3}) is an asymptotically optimal QCSS with parameters
\[
((\mu_{\min}-1)N,N-1,N,N).
\]
\end{thm}
\begin{IEEEproof}
It is immediate that $\widehat{\mathfrak{C}}$ has flock size $N-1$ and sequence length $N$. Notice that
\begin{align*}
R_{\widehat{\cc}^{a_1,b_1}, \widehat{\mathfrak{C}}^{a_2,b_2}}(\tau)
&= \sum_{i=1}^{N-1} \sum_{j=0}^{N-1} \xi_N^{(a_1\rho(i)+b_1)j-(a_2\rho(i)+b_2)(j+\tau)}\\
&= \xi_N^{-b_2\tau} \sum_{i=1}^{N-1}\xi_N^{-a_2\rho(i)\tau} \sum_{j=0}^{N-1}\xi_N^{\left((a_1-a_2) \rho(i)+b_1-b_2\right)j}.
\end{align*}
As before, we cover all possible cases to determine $\left|R_{\widehat{\mathfrak{C}}^{a_1,b_1}, \widehat{\mathfrak{C}}^{a_2,b_2}}(\tau)\right|$.

\begin{enumerate}[wide, itemsep=0pt, leftmargin =0pt,widest={{\bf Case $2$}}]

\item[{\bf Case $1$}:] If $a_1=a_2$, $b_1=b_2$, and $\tau\in[N]\setminus\{0\}$, then
\[
R_{\widehat{\mathfrak{C}}^{a_1,b_1},\widehat{\mathfrak{C}}^{a_2,b_2}}(\tau)=N\xi_N^{-b_2\tau}\sum_{i=1}^{N-1}\xi_N^{-a_2\rho(i)\tau}.
\]
Since $\tau\in[N]\setminus\{0\}$ and $a_2\in [\mu_{\min}]\setminus\{0\}$, we have $a_2\tau\not\equiv 0 \pmod N$, which implies
\[
\left|R_{\widehat{\mathfrak{C}}^{a_1,b_1}, \widehat{\mathfrak{C}}^{a_2,b_2}}(\tau)\right| = \left|-N\xi_N^{-b_2\tau}\right| = N.
\]

\item[{\bf Case $2$}:]If $a_1=a_2$, $b_1\neq b_2$, and $\tau\in[N]$, then
\[
R_{\widehat{\mathfrak{C}}^{a_1,b_1},\widehat{\mathfrak{C}}^{a_2,b_2}}(\tau)=\xi_N^{-b_2\tau}\sum_{i=1}^{N-1}\xi_N^{-a_2\rho(i)\tau}\sum_{j=0}^{N-1}\xi_N^{\left(b_1-b_2\right)j}=0.
\]

\item[{\bf Case $3$}:]If $a_1\neq a_2$, $b_1=b_2$, and $\tau\in[N]$, then it follows from the fact that $(a_1-a_2)\rho(i)\not\equiv 0 \pmod N$ for each $i \in[N]\setminus\{0\}$ that
\[
R_{\widehat{\mathfrak{C}}^{a_1,b_1}, \widehat{\mathfrak{C}}^{a_2,b_2}}(\tau) = \xi_N^{-b_2\tau} \sum_{i=1}^{N-1} \xi_N^{-a_2\rho(i)\tau} \sum_{j=0}^{N-1}\xi_N^{\left((a_1-a_2)\rho(i)\right)j} = 0.
\]

\item[{\bf Case $4$}:]If $a_1\neq a_2$, $b_1\neq b_2$, and $\tau\in[N]$, then
\begin{eqnarray*}
R_{\widehat{\mathfrak{C}}^{a_1,b_1},\widehat{\mathfrak{C}}^{a_2,b_2}}(\tau)=\xi_N^{-b_2\tau}\sum_{i=1}^{N-1}\xi_N^{-a_2\rho(i)\tau}\sum_{j=0}^{N-1}\xi_N^{\left((a_1-a_2)\rho(i)+b_1-b_2\right)j}.
\end{eqnarray*}
By Lemma \ref{lem41}, the congruence $(a_1-a_2)\rho(i) \equiv b_2-b_1 \pmod N$ has a unique solution $\ell$, since $\gcd(a_1-a_2,N)=1$. Hence,
\begin{align*}
\left|R_{\widehat{\mathfrak{C}}^{a_1,b_1},\widehat{\mathfrak{C}}^{a_2,b_2}}(\tau)\right|
&= \left|\xi_N^{-b_2\tau}\xi_N^{-a_2\rho(\ell)\tau}\sum_{j=0}^{N-1}1+\xi_N^{-b_2\tau}\sum_{i=1,i\neq \ell}^{N-1}\xi_N^{-a_2\rho(i)\tau}\sum_{j=0}^{N-1}\xi_N^{\left((a_1-a_2)\rho(i)+b_1-b_2\right)j}\right|\\
&= \left|N\xi_N^{-b_2\tau}\xi_N^{-a_2\rho(\ell)\tau}\right| = N.
\end{align*}
\end{enumerate}
Considering all four cases, $\widehat{\mathfrak{C}}$ has $\vartheta_{\max} = N$. Using the same argument as in the proof of Theorem \ref{thm41}, we can show that $\widehat{\mathfrak{C}}$ has family size $M=(\mu_{\min}-1)N$ and asymptotically attains the bound in (\ref{bound}).
\end{IEEEproof}

\begin{remark}
The construction of QCSSs in Theorem \ref{thm42} is derived from that of Theorem \ref{thm41} by deleting the first row in each complementary sequence and setting $\rho$ to be any permutation that fixes $0$. An asymptotically optimal $((\mu_{\min}-1)N,N-1,N,N)$-QCSS can in fact be produced by deleting the $i^{{\rm th}}$ row in each element of the QCSS constructed in Theorem \ref{thm41}, where $1 \leq i \leq N$. The permutation $\rho$ can be any permutation. The computation of the parameters of the resulting QCSS is analogous to the one carried out in the proof of Theorem \ref{thm42}.
\end{remark}

The next example shows how a periodic QCSS can be explicitly built based on Theorem \ref{thm42}.

\begin{example}
Let $N=9$, $\rho(x)=x$, and $\xi_{9}:=e^{2\pi\sqrt{-1}/9}$. The $\widehat{\mathfrak{C}}$ defined by Theorem \ref{thm42} is a periodic $(18,8,9,9)$-QCSS consisting of the following complementary sequences, with every original entry of $\cc^i$ replaced by the corresponding exponent when expressed as a power of $\xi_{9}$.
\[
\cc^0=\begin{pmatrix}
\begin{smallmatrix}
  0& 1& 2& 3& 4& 5& 6& 7& 8  \\
  0& 2& 4& 6& 8& 1& 3& 5& 7  \\
  0& 3& 6& 0& 3& 6& 0& 3& 6  \\
  0& 4& 8& 3& 7& 2& 6& 1& 5  \\
  0& 5& 1& 6& 2& 7& 3& 8& 4  \\
  0& 6& 3& 0& 6& 3& 0& 6& 3  \\
  0& 7& 5& 3& 1& 8& 6& 4& 2  \\
  0& 8& 7& 6& 5& 4& 3& 2& 1
\end{smallmatrix}
\end{pmatrix}, \quad
\cc^1=\begin{pmatrix}
\begin{smallmatrix}
  0& 2& 4& 6& 8& 1& 3& 5& 7  \\
  0& 4& 8& 3& 7& 2& 6& 1& 5  \\
  0& 6& 3& 0& 6& 3& 0& 6& 3  \\
  0& 8& 7& 6& 5& 4& 3& 2& 1  \\
  0& 1& 2& 3& 4& 5& 6& 7& 8  \\
  0& 3& 6& 0& 3& 6& 0& 3& 6  \\
  0& 5& 1& 6& 2& 7& 3& 8& 4  \\
  0& 7& 5& 3& 1& 8& 6& 4& 2
\end{smallmatrix}
\end{pmatrix}, \quad
\cc^2=\begin{pmatrix}
\begin{smallmatrix}
  0& 2& 4& 6& 8& 1& 3& 5& 7  \\
  0& 3& 6& 0& 3& 6& 0& 3& 6  \\
  0& 4& 8& 3& 7& 2& 6& 1& 5  \\
  0& 5& 1& 6& 2& 7& 3& 8& 4  \\
  0& 6& 3& 0& 6& 3& 0& 6& 3  \\
  0& 7& 5& 3& 1& 8& 6& 4& 2  \\
  0& 8& 7& 6& 5& 4& 3& 2& 1  \\
  0& 0& 0& 0& 0& 0& 0& 0& 0
\end{smallmatrix}
\end{pmatrix},
\]
\[
\cc^3=\begin{pmatrix}
\begin{smallmatrix}
  0& 3& 6& 0& 3& 6& 0& 3& 6  \\
  0& 5& 1& 6& 2& 7& 3& 8& 4  \\
  0& 7& 5& 3& 1& 8& 6& 4& 2  \\
  0& 0& 0& 0& 0& 0& 0& 0& 0  \\
  0& 2& 4& 6& 8& 1& 3& 5& 7  \\
  0& 4& 8& 3& 7& 2& 6& 1& 5  \\
  0& 6& 3& 0& 6& 3& 0& 6& 3  \\
  0& 8& 7& 6& 5& 4& 3& 2& 1
\end{smallmatrix}
\end{pmatrix}, \quad
\cc^4=\begin{pmatrix}
\begin{smallmatrix}
  0& 3& 6& 0& 3& 6& 0& 3& 6  \\
  0& 4& 8& 3& 7& 2& 6& 1& 5  \\
  0& 5& 1& 6& 2& 7& 3& 8& 4  \\
  0& 6& 3& 0& 6& 3& 0& 6& 3  \\
  0& 7& 5& 3& 1& 8& 6& 4& 2  \\
  0& 8& 7& 6& 5& 4& 3& 2& 1  \\
  0& 0& 0& 0& 0& 0& 0& 0& 0  \\
  0& 1& 2& 3& 4& 5& 6& 7& 8
\end{smallmatrix}
\end{pmatrix}, \quad
\cc^5=\begin{pmatrix}
\begin{smallmatrix}
  0& 4& 8& 3& 7& 2& 6& 1& 5  \\
  0& 6& 3& 0& 6& 3& 0& 6& 3  \\
  0& 8& 7& 6& 5& 4& 3& 2& 1  \\
  0& 1& 2& 3& 4& 5& 6& 7& 8  \\
  0& 3& 6& 0& 3& 6& 0& 3& 6  \\
  0& 5& 1& 6& 2& 7& 3& 8& 4  \\
  0& 7& 5& 3& 1& 8& 6& 4& 2  \\
  0& 0& 0& 0& 0& 0& 0& 0& 0
\end{smallmatrix}
\end{pmatrix},
\]
\[
\cc^6=\begin{pmatrix}
\begin{smallmatrix}
  0& 4& 8& 3& 7& 2& 6& 1& 5  \\
  0& 5& 1& 6& 2& 7& 3& 8& 4  \\
  0& 6& 3& 0& 6& 3& 0& 6& 3  \\
  0& 7& 5& 3& 1& 8& 6& 4& 2  \\
  0& 8& 7& 6& 5& 4& 3& 2& 1  \\
  0& 0& 0& 0& 0& 0& 0& 0& 0  \\
  0& 1& 2& 3& 4& 5& 6& 7& 8  \\
  0& 2& 4& 6& 8& 1& 3& 5& 7
\end{smallmatrix}
\end{pmatrix}, \quad
\cc^7=\begin{pmatrix}
\begin{smallmatrix}
  0& 5& 1& 6& 2& 7& 3& 8& 4  \\
  0& 7& 5& 3& 1& 8& 6& 4& 2  \\
  0& 0& 0& 0& 0& 0& 0& 0& 0  \\
  0& 2& 4& 6& 8& 1& 3& 5& 7  \\
  0& 4& 8& 3& 7& 2& 6& 1& 5  \\
  0& 6& 3& 0& 6& 3& 0& 6& 3  \\
  0& 8& 7& 6& 5& 4& 3& 2& 1  \\
  0& 1& 2& 3& 4& 5& 6& 7& 8
\end{smallmatrix}
\end{pmatrix}, \quad
\cc^8=\begin{pmatrix}
\begin{smallmatrix}
  0& 5& 1& 6& 2& 7& 3& 8& 4  \\
  0& 6& 3& 0& 6& 3& 0& 6& 3  \\
  0& 7& 5& 3& 1& 8& 6& 4& 2  \\
  0& 8& 7& 6& 5& 4& 3& 2& 1  \\
  0& 0& 0& 0& 0& 0& 0& 0& 0  \\
  0& 1& 2& 3& 4& 5& 6& 7& 8  \\
  0& 2& 4& 6& 8& 1& 3& 5& 7  \\
  0& 3& 6& 0& 3& 6& 0& 3& 6
\end{smallmatrix}
\end{pmatrix},
\]
\[
\cc^9=\begin{pmatrix}
\begin{smallmatrix}
  0& 6& 3& 0& 6& 3& 0& 6& 3  \\
  0& 8& 7& 6& 5& 4& 3& 2& 1  \\
  0& 1& 2& 3& 4& 5& 6& 7& 8  \\
  0& 3& 6& 0& 3& 6& 0& 3& 6  \\
  0& 5& 1& 6& 2& 7& 3& 8& 4  \\
  0& 7& 5& 3& 1& 8& 6& 4& 2  \\
  0& 0& 0& 0& 0& 0& 0& 0& 0  \\
  0& 2& 4& 6& 8& 1& 3& 5& 7
\end{smallmatrix}
\end{pmatrix}, \quad
\cc^{10}=\begin{pmatrix}
\begin{smallmatrix}
  0& 6& 3& 0& 6& 3& 0& 6& 3  \\
  0& 7& 5& 3& 1& 8& 6& 4& 2  \\
  0& 8& 7& 6& 5& 4& 3& 2& 1  \\
  0& 0& 0& 0& 0& 0& 0& 0& 0  \\
  0& 1& 2& 3& 4& 5& 6& 7& 8  \\
  0& 2& 4& 6& 8& 1& 3& 5& 7  \\
  0& 3& 6& 0& 3& 6& 0& 3& 6  \\
  0& 4& 8& 3& 7& 2& 6& 1& 5
\end{smallmatrix}
\end{pmatrix}, \quad
\cc^{11}=\begin{pmatrix}
\begin{smallmatrix}
  0& 7& 5& 3& 1& 8& 6& 4& 2  \\
  0& 0& 0& 0& 0& 0& 0& 0& 0  \\
  0& 2& 4& 6& 8& 1& 3& 5& 7  \\
  0& 4& 8& 3& 7& 2& 6& 1& 5  \\
  0& 6& 3& 0& 6& 3& 0& 6& 3  \\
  0& 8& 7& 6& 5& 4& 3& 2& 1  \\
  0& 1& 2& 3& 4& 5& 6& 7& 8  \\
  0& 3& 6& 0& 3& 6& 0& 3& 6
\end{smallmatrix}
\end{pmatrix},
\]
\[
\cc^{12}=\begin{pmatrix}
\begin{smallmatrix}
  0& 7& 5& 3& 1& 8& 6& 4& 2  \\
  0& 8& 7& 6& 5& 4& 3& 2& 1  \\
  0& 0& 0& 0& 0& 0& 0& 0& 0  \\
  0& 1& 2& 3& 4& 5& 6& 7& 8  \\
  0& 2& 4& 6& 8& 1& 3& 5& 7  \\
  0& 3& 6& 0& 3& 6& 0& 3& 6  \\
  0& 4& 8& 3& 7& 2& 6& 1& 5  \\
  0& 5& 1& 6& 2& 7& 3& 8& 4
\end{smallmatrix}
\end{pmatrix}, \quad
\cc^{13}=\begin{pmatrix}
\begin{smallmatrix}
  0& 8& 7& 6& 5& 4& 3& 2& 1  \\
  0& 1& 2& 3& 4& 5& 6& 7& 8  \\
  0& 3& 6& 0& 3& 6& 0& 3& 6  \\
  0& 5& 1& 6& 2& 7& 3& 8& 4  \\
  0& 7& 5& 3& 1& 8& 6& 4& 2  \\
  0& 0& 0& 0& 0& 0& 0& 0& 0  \\
  0& 2& 4& 6& 8& 1& 3& 5& 7  \\
  0& 4& 8& 3& 7& 2& 6& 1& 5
\end{smallmatrix}
\end{pmatrix}, \quad
\cc^{14}=\begin{pmatrix}
\begin{smallmatrix}
  0& 8& 7& 6& 5& 4& 3& 2& 1  \\
  0& 0& 0& 0& 0& 0& 0& 0& 0  \\
  0& 1& 2& 3& 4& 5& 6& 7& 8  \\
  0& 2& 4& 6& 8& 1& 3& 5& 7  \\
  0& 3& 6& 0& 3& 6& 0& 3& 6  \\
  0& 4& 8& 3& 7& 2& 6& 1& 5  \\
  0& 5& 1& 6& 2& 7& 3& 8& 4  \\
  0& 6& 3& 0& 6& 3& 0& 6& 3
\end{smallmatrix}
\end{pmatrix},
\]
\[
\cc^{15}=\begin{pmatrix}
\begin{smallmatrix}
  0& 0& 0& 0& 0& 0& 0& 0& 0  \\
  0& 2& 4& 6& 8& 1& 3& 5& 7  \\
  0& 4& 8& 3& 7& 2& 6& 1& 5  \\
  0& 6& 3& 0& 6& 3& 0& 6& 3  \\
  0& 8& 7& 6& 5& 4& 3& 2& 1  \\
  0& 1& 2& 3& 4& 5& 6& 7& 8  \\
  0& 3& 6& 0& 3& 6& 0& 3& 6  \\
  0& 5& 1& 6& 2& 7& 3& 8& 4
\end{smallmatrix}
\end{pmatrix}, \quad
\cc^{16}=\begin{pmatrix}
\begin{smallmatrix}
  0& 0& 0& 0& 0& 0& 0& 0& 0  \\
  0& 1& 2& 3& 4& 5& 6& 7& 8  \\
  0& 2& 4& 6& 8& 1& 3& 5& 7  \\
  0& 3& 6& 0& 3& 6& 0& 3& 6  \\
  0& 4& 8& 3& 7& 2& 6& 1& 5  \\
  0& 5& 1& 6& 2& 7& 3& 8& 4  \\
  0& 6& 3& 0& 6& 3& 0& 6& 3  \\
  0& 7& 5& 3& 1& 8& 6& 4& 2
\end{smallmatrix}
\end{pmatrix}, \quad
\cc^{17}=\begin{pmatrix}
\begin{smallmatrix}
  0& 1& 2& 3& 4& 5& 6& 7& 8  \\
  0& 3& 6& 0& 3& 6& 0& 3& 6  \\
  0& 5& 1& 6& 2& 7& 3& 8& 4  \\
  0& 7& 5& 3& 1& 8& 6& 4& 2  \\
  0& 0& 0& 0& 0& 0& 0& 0& 0  \\
  0& 2& 4& 6& 8& 1& 3& 5& 7  \\
  0& 4& 8& 3& 7& 2& 6& 1& 5  \\
  0& 6& 3& 0& 6& 3& 0& 6& 3
\end{smallmatrix}
\end{pmatrix}.
\]
\QEDB
\end{example}

\section{Concluding Remarks}\label{S6}

As the number of connected devices continues to grow, often at an exponential rate, the search for better performing schemes follows suit. Our work in this paper takes inspiration from practical needs to design efficient and reliable sets of sequences that possess excellent properties. A great set for the purpose must not only have a large size for given flock size and length but also the maximum possible amount of multipath and multiuser interference.

We have used the interleaving approach to come up with a general construction of quasi-complementary sequence sets. The constructed sequences are asymptotically optimal. The general construction allows for seven infinite families of such sequences to be built explicitly. The insights gained then lead us to two further constructions. The resulting two families of sequence sets are asymptotically optimal. More attractive are their flexible parameters and smaller alphabet sizes. These favourable attributes give system designers refined tools when faced with specific deployment requirements and constraints.

The main mathematical tools come from the theory of groups in the form of additive and multiplicative characters of finite fields. Their elegance and simplicity make the proposed constructions easy to understand and straightforward to implement.

\end{document}